\def\year{2020}\relax
\documentclass[letterpaper]{article} 
\usepackage{aaai20}  
\usepackage{times}  
\usepackage{helvet} 
\usepackage{courier}  
\usepackage[hyphens]{url}  
\usepackage{graphicx} 
\usepackage{graphicx}  
\usepackage{subfig}
\usepackage{xspace}

 \usepackage{xcolor}

\newcommand{\one}{({\em i}\/)\xspace}
\newcommand{\two}{({\em ii}\/)\xspace}

\def\eg{\emph{e.g., }\xspace}
\def\etc{\emph{etc.}\xspace}
\def\ie{\emph{i.e.,}\xspace}
\def\etal{\emph{et al.}\xspace}

\begin{document}

\title{Characterising User Content on a Multi-lingual Social Network}
\author{Pushkal Agarwal,\textsuperscript{\rm 1}
Kiran Garimella,\textsuperscript{\rm 2}
Sagar Joglekar,\textsuperscript{\rm 1}
Nishanth Sastry,\textsuperscript{\rm 1,4}
Gareth Tyson,\textsuperscript{\rm 3,4}\\
\textsuperscript{\rm 1}King's College London,
\textsuperscript{\rm 2}MIT,
\textsuperscript{\rm 3}Queen Mary University of London,
\textsuperscript{\rm 4}Alan Turing Institute\\
pushkal.agarwal@kcl.ac.uk, garimell@mit.edu, sagar.joglekar@kcl.ac.uk,
nishanth.sastry@kcl.ac.uk, g.tyson@qmul.ac.uk
}

\maketitle

\begin{abstract}
Social media has been on the vanguard of political information diffusion in the 21st century. Most studies that look into disinformation, political influence and fake-news focus on mainstream social media platforms. This has inevitably made English an important factor in our current understanding of political activity on social media. 
As a result, there has only been a limited number of studies into a large portion of the world, including the largest, multilingual and multicultural democracy: India. 
In this paper we present our characterisation of a multilingual social network in India called ShareChat. We collect an exhaustive dataset across 72 weeks before and during the Indian general elections of 2019, across 14 languages. We investigate the cross lingual dynamics by clustering visually similar images together, and exploring how they move across language barriers.
We find that Telugu, Malayalam, Tamil and Kannada languages tend to be dominant in soliciting political images (often referred to as memes), and posts from Hindi have the largest cross-lingual diffusion across ShareChat (as well as images containing text in English). 
In the case of images containing text that cross language barriers, we see that language translation is used to widen the accessibility. 
That said, we find cases where the same image is associated with very different text (and therefore meanings).
This initial characterisation paves the way for more advanced pipelines to understand the dynamics of fake and political content in a multi-lingual and non-textual setting.

\end{abstract}


\section{Introduction}

Soon after its independence, India was divided internally into states, based  mainly on the different languages spoken in each region, thus creating a distinct regional identity among its citizens in addition to the idea of nationhood~\cite{guha2017india}. There have often been conflicts and disagreements across state borders, where separate regional identities have been asserted over national unity~\cite{gupta1970language}. 
Within this historical and political context, we wish to understand the extent to which information is shared across different languages. 
 
We take as our case study the 2019 Indian General Election, considered to be the largest ever exercise in representative democracy, with over 67\% of the 900 million strong electorate participating in the polls. 
Similar to recent elections in other countries~\cite{metaxas2012social,agarwal2019tweeting}, it is also undeniable that social media played an important role in the Indian General Elections, helping mobilise voters and spreading information about the different major parties~\cite{rao2019role,patel2019indian}. 
Given the rich linguistic diversity in India, we wish to understand the manner in which such a large scale \emph{national} effort plays out on social media despite the differences in languages among the electorate.
We focus our efforts on understanding whether and to what extent information  on social media crosses regional and linguistic divides.

To explore this, we have collected the first large-scale dataset, consisting of over 1.2 million posts across 14 languages 
during and before the election campaigning period, from \textbf{ShareChat}.\footnote{An anonymised version of the dataset is available for non-commercial research usage from \url{https://tiny.cc/share-chat}.} This is a media and text-sharing platform designed for Indian users, with over 50 million monthly active users.\footnote{\url{https://we.sharechat.com/}}
In functionality, it is similar to services like Instagram, with users sharing mostly multimedia content. However, it has one major difference that aids our research design:  Unlike other global platforms, different languages are separated into  communities of their own, which creates a silo effect and a strong sense of commonality.
Despite this regional appeal, it has accumulated tens of millions of users in just 4 years of existence, with most being first time Internet users~\cite{yourstory}. 

\begin{figure}
\centering
  \includegraphics[width=\columnwidth]{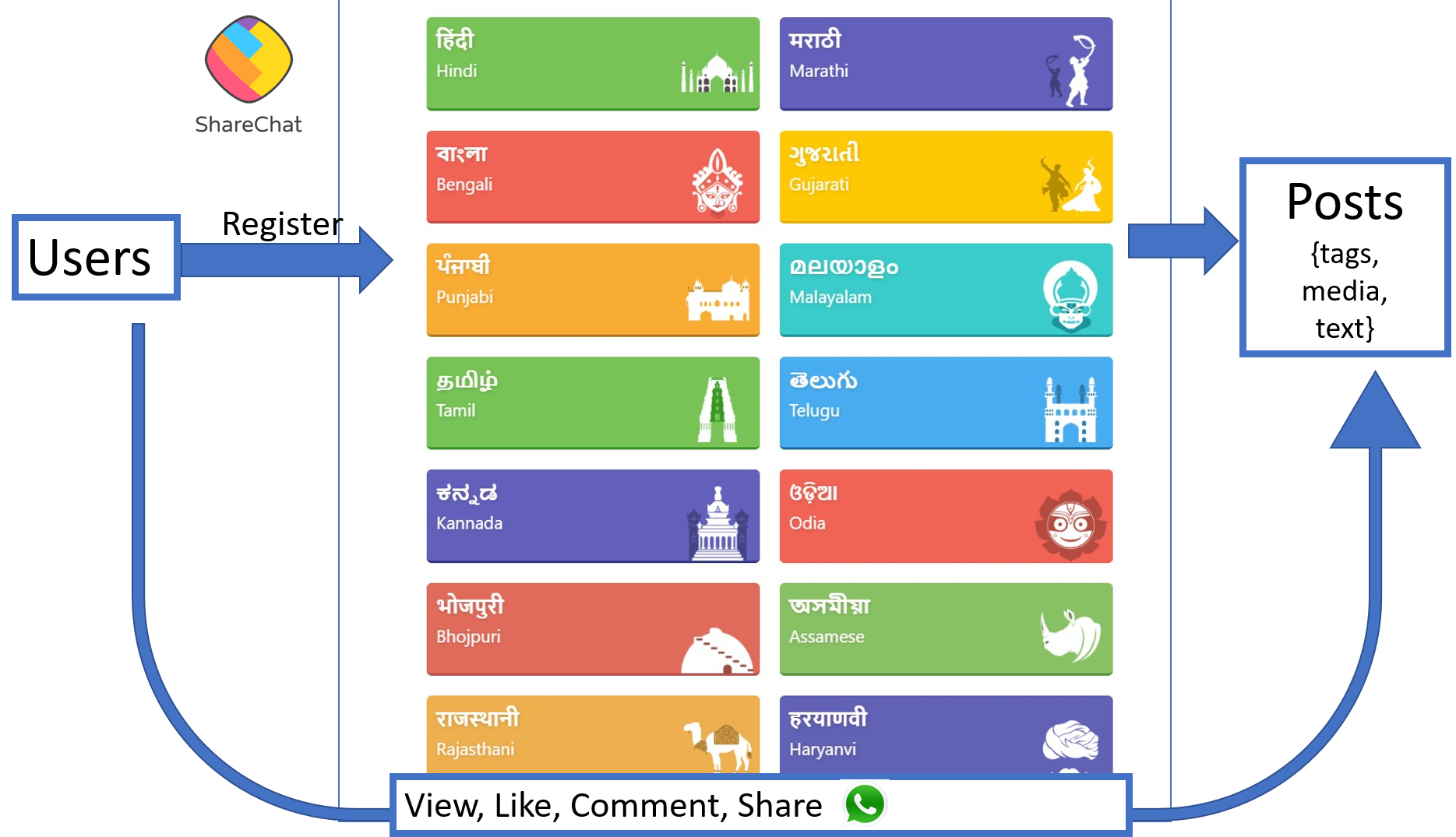}
  \caption{ShareChat homepage and 
  user interactions.}
  \label{fig:appPage}
\end{figure}

Figure~\ref{fig:appPage} presents the ShareChat homepage, in which users must first select the language community into which they post. Note that there is \emph{no} support for English language during the registration process,\footnote{\url{https://techcrunch.com/2019/08/15/sharechat-twitter-seriesd/}} thereby creating a unique social media platform for regional languages. 
Thus, by mirroring the language-based divisions of the political map of India, ShareChat offers a fascinating environment to study Indian identity along national and regional lines. This is in contrast to other relevant platforms, such as WhatsApp, which do not have such formal language boundaries~\cite{melo2019whatsapp,rao2019role,garimella2018whatapp}.

Our key hypothesis is that these language barriers may be overcome via image-based content, which is less dependent on linguistic understanding than text or audio. To explore this, we formulate the following two research questions:

\begin{description}
    \item[RQ-1] Can image content transcend language silos? If so, how often does this happen, and amongst which languages?
    
    \item[RQ-2] What kinds of images are most successful in transcending language silos, and do their semantic meanings mutate?
    
\end{description}

To answer these questions, we exploit our ShareChat data and propose a methodology to cluster perceptually similar images, across languages, and understand how they change per community.
We find that a number of these images are contain associated text~\cite{zannettou2018origins}. We therefore extract language features within images via Optical Character Recognition (OCR) and further annotate multi-lingual images with translations and semantic tags. 

Using this data, we investigate the characteristics of images that have a wider cross-lingual adoption. We find that a majority of the images have embedded text in them, which is recognised by OCR. This presence of text strongly affects the sharing of images across languages. We find, for example, that sharing is easier when the image text is in languages such as Hindi and English, which are lingua franca widely understood across large portions of India. We also find that the text of the image is translated when shared across different language communities, and that images spread more easily across linguistically related languages or in languages spoken in adjacent regions. We even observe that sometimes message change during translation, \eg political memes are altered to make them more consumable in a specific language, or the meaning is altered on purpose in the shared text. Our results have key implications for understanding political communications in multi-lingual environments. 

\section{Related Work}

Our study focuses on multi-lingual political discourse in the largest democracy in the world, India. 
There have been a number of works looking into the multi-lingual nature of certain platforms.
For example, language differences amongst Wikipedia editors~\cite{kaffee2018analysis}, as well as general differences amongst language editions~\cite{hale2014multilinguals,hecht2010tower,hale2015cross}. Often this goes beyond textual differences, showing that even images differ amongst language editions~\cite{he2018the_tower_of_babel}.
There have also been efforts to bridge these gaps~\cite{bao2012omnipedia}, although often differences extend beyond language to include differences between facts and foci. 
This often leads editors to work on their non-native language editions~\cite{hale2015cross}. Although highly relevant, our work differs in that we focus on a social community platform, in which individuals and groups communicate. This contrasts starkly with the use of community-driven online encyclopedias, which are focused on communal curation of knowledge.

More closely related are the range of studies looking at the multi-lingual use of social platforms, \eg blogs~\cite{hale2012net}, reviews~\cite{hale2016user} and Twitter~\cite{nguyen2015audience,hong2011language}. 
In most cases, these studies find a dominance of English language material. For example, Hong \etal~\cite{hong2011language} found that 51\% of tweets are in English. 
Curiously, it has been found that individuals often change their language choices to reflect different positions or opinions~\cite{murthy2015we}. 
Another major part of this paper is the study of memes and image sharing. Recent works have looked at meme sharing on platforms like Reddit, 4Chan and Gab, where they are often used to spread hate and fake news~\cite{zannettou2018origins}. These modalities are powerful, in that they often transcend languages, and have sometimes provoked real world consequences~\cite{zannettou2019let} or have been used as a mode of information warfare~\cite{savvasDisinfo,raman2019challenges}. 


Our research differs substantially from the works above. The above social media platform does not embed language directly into their operations. Instead, language communities form naturally. In contrast, ShareChat is \emph{themed} around language with strict distinctions between communities. Furthermore, it has a strong focus on image-based sharing, which we posit may not be impacted by language in the same way as textual posts. As such, we are curious to understand the interplay between these strictly defined language communities, and the capacity of images to cross language boundaries. To the best of our knowledge, this is the first paper to empirically explore user and language behaviour at such scale on a multimedia heavy, multi-lingual platform, across languages used (in India).

\section{Data Collection}


We have gathered, to the best of our knowledge, the first large-scale multi-lingual social network dataset. To collect this dataset, we started with the set of manually curated topics that Sharechat publishes everyday on their website, for each language.\footnote{\url{https://sharechat.com/explore}} These topics are separated across various categories, such as entertainment, politics, religion, news, with each topic containing a set of popular hashtags.
In this paper, we focus on politics and therefore gather all  hashtags posted about politics between 28 January 2018 and 20 June 2019, across all the 14 languages supported by ShareChat. In total, 1,313 hashtags were used, of which 480 were related to Politics. Thus, we  believe Politics represents a significant portion of activity on Sharechat during the period of study. Note that the period we consider coincides with high profile events of national importance, such as the Indian National Elections and an escalation of the India-Pakistan conflict.

We obtained all the posts from all the political hashtags using the ShareChat API, for the entire duration of the crawl. Each post is serialised as a JSON object containing the complete metadata about the post, including the number of likes, comments, shares on WhatsApp,\footnote{Unlike Twitter, ShareChat does not have a retweet button, but allows users to quickly share content to WhatsApp. Hence the share counts reported here are shares from ShareChat to WhatsApp.} tags, post identifier, Optical Character Recognition (OCR) text from the image and date of creation.

We further download all the images, video and audio content from the posts.
Overall, our dataset consists of 1.2 million posts from 321k unique users.
These posts consist of a diverse set of media types (Gifs, images, videos), out of which almost half (641k) are images.
Since images are the most dominant medium of communication, in the following sections, we only consider posts containing images. Images also receive significantly more engagement on the platform, with a median of \textbf{377} views per image as opposed to \textbf{172} for non-images.

Around 15\% of images have no OCR text (i.e. no textual content in the image) in them. 
We manually check the accuracy of  OCR on a few hundred posts in multiple languages and find a satisfactory performance (Figure~\ref{fig:qualitative}). 
We also identify the language from the OCR text and hashtags using Google's Compact Language Detector~\cite{cld3}. 
In the rest of the paper, the term \emph{language} refers to the profile language of the user who authored a post. When referencing languages identified from post processing, we will precede it with the method used to identify the language, i.e., \emph{OCR language} or \emph{Tag language}.
We believe that our large scale multi-lingual, multi-modal dataset would be  valuable for research in Computational Social Science, and could also benefit researchers from other fields including Natural Language Processing, and Computer Vision.

\section{Data Processing Methodology}

We next process the data to \one~cluster similar images together; and \two~translate all text into English. 

\subsection{Image Clustering} \label{sec:img-cluster}
Since we are dealing with more than half a million images, in order to make the analysis tractable, we cluster together visually and perceptually similar images. 
This allows us to effectively find ``memes''~\cite{zannettou2018origins}. For simplicity, we refer to any images containing accompanying text as memes.
To identify clusters of images, we make use of a recently open sourced image hashing algorithm by Facebook known as PDQ.\footnote{github.com/facebook/ThreatExchange/tree/master/hashing/pdq}
The PDQ hashing algorithm is a significant improvement over the commonly used phash~\cite{zauner2010implementation}, and produces a 256 bit hash using a discrete cosine transformation algorithm. 
PDQ is used by Facebook to detect similar content, and is the  state-of-the-art approach for identifying and clustering similar images.

Using PDQ, we cluster images and inspect each cluster for features they correspond to. The clustering takes one parameter $d$: the distance threshold for the images to be considered similar. This parameter takes values in the range 31--63. Through manual inspection, we find that for $d>50$,  images that are very different from each other tend to get grouped in the same cluster, and for $d<50$, the choice of $d$ does not appear to make much of a difference, and clusters typically show a high degree of cohesion (i.e., the clusters largely consist of copies of one unique image).  Therefore, having tested multiple possible values of the distance parameter, we choose $d=31$, the default value used in PDQ, as this yields cohesive clusters.
Through this process, we obtain over 400k image clusters from 560k individual images. Out of these, 54k clusters have between 2--10 images, and roughly 2000 clusters have over 10 images.
The biggest cluster contains 261 images.


\subsection{OCR Language Translation}
\label{sec:ocr}
To enable analysis, we next translate all the OCR text contained within images into English using Google Translate. 
To validate correctness, we ask native language speakers from 8 of the 10 languages we consider, to verify the translations.
Our validation exercise focuses on 63 clusters (containing 769 images) which are identified as having images with more than 3 different languages, according to OCR. The native speakers first check whether the OCR has correctly identified the text contained in the image, and then check if the machine translation by Google Translate is accurate. In the case of errors in either of these two steps, our volunteers provided high quality manual translations to provide an accurate ground truth.

\begin{figure}
    \centering
    \includegraphics[width=\columnwidth]{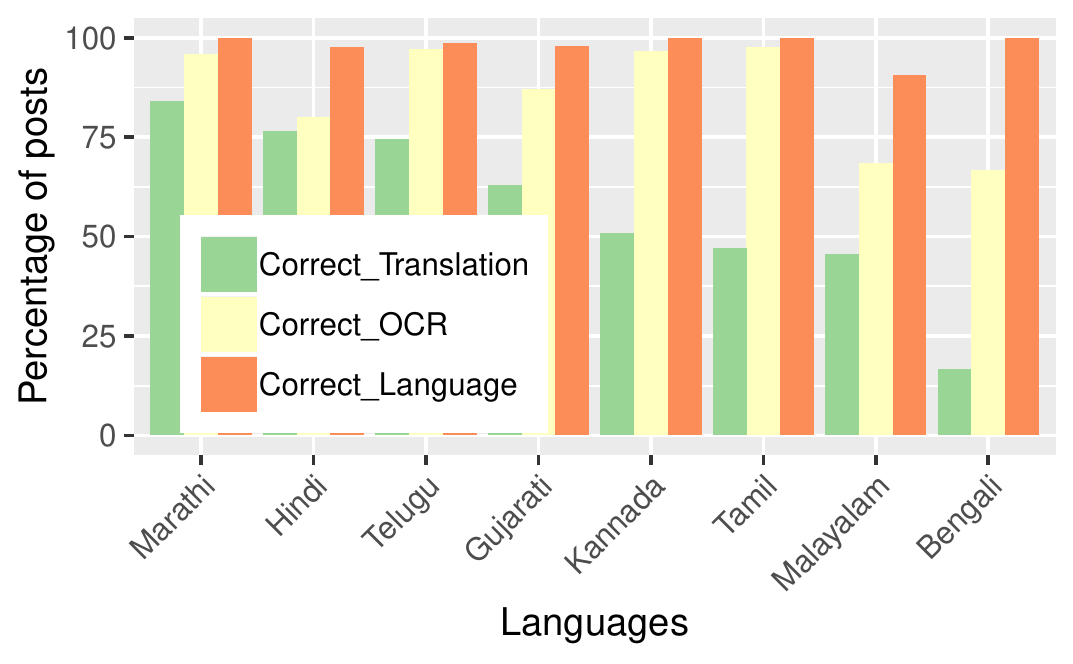}
    \caption{Accuracy of translation and language detection, as evaluated from a sample of 769 images by native speakers: Orange bar shows the percentage of posts for which the language detected by OCR was accurate (close to 100\% for all languages). Yellow bar shows the accuracy of the OCR-generated text ($>75$\% for all languages except Malayalam and Bengali). Green bar shows the accuracy of automatic translation into English.}
    \label{fig:qualitative}
\end{figure}

In Figure~\ref{fig:qualitative} we first present the recorded accuracy of the translation of OCR from the native language to English (orange bar).
Overall, this shows that the OCR language detection used by ShareChat performs well (90\% to 100\%).
Figure~\ref{fig:qualitative} also shows the accuracy of the extracted OCR text (yellow bar). Based on the language, we see mostly positive results (\ie above 90\% accuracy), except for a few, \eg Bengali which falls under 70\%.
Finally, the figure shows that automatic translation by Google is not that accurate (green bar). 
For certain widely spoken languages such as Hindi, translation performs well: over 75\% of translations require no modification. 
However, this does not hold true for many other Indian languages. 
For example, with Telugu, translation was sensitive to spaces, and performed poorly for long sentences. In contrast, Tamil translations prove to be poorer for short sentences. Furthermore, even minor typographical mistakes in the OCR text negatively impact the results, without the capacity to `auto correct' as seen with English.

\section{Basic Characterisation}

We begin by providing a brief statistical overview of activity on ShareChat. 

\subsection{Summary Statistics}

\begin{figure}
\centering
  \includegraphics[width=0.9\columnwidth]{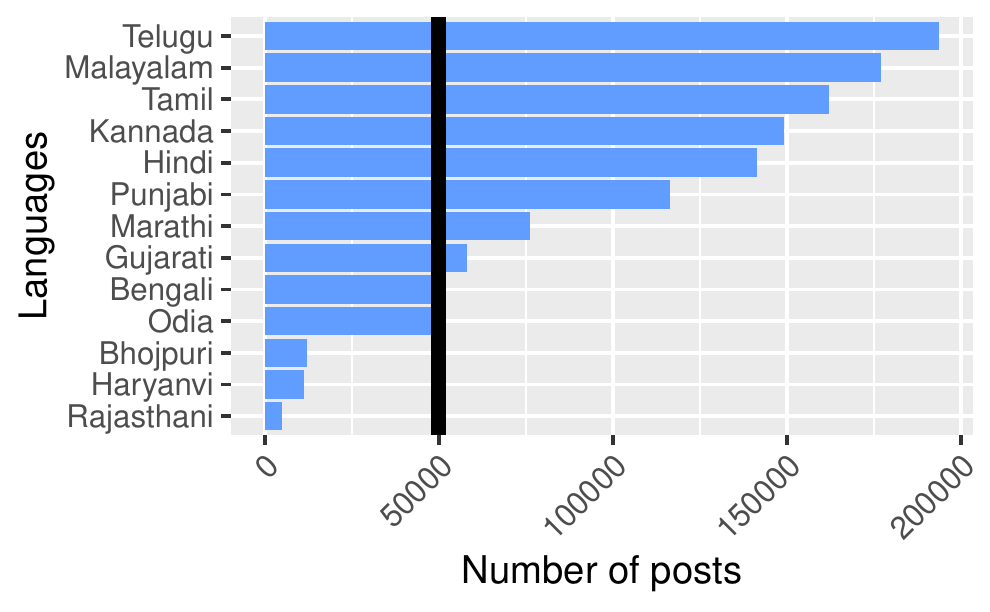}
  \caption{Posts count per language.}
  \label{fig:postCountLang}
\end{figure}

Figure~\ref{fig:postCountLang} presents the number of posts in each language.
Due the varying population sizes, we see a strong skew towards widely spoken languages. Surprisingly, Hindi is \emph{not} the most popular language though. Instead, Telugu and Malayalam accumulate the majority of posts (16.4\% and 15\% respectively). Due to this skew, in the later sections we choose to use just the top 10 languages, as languages such as  Bhojpuri, Haryanvi and Rajasthani accumulate very few posts. 

\begin{figure}
\centering
  \includegraphics[width=0.9\columnwidth]{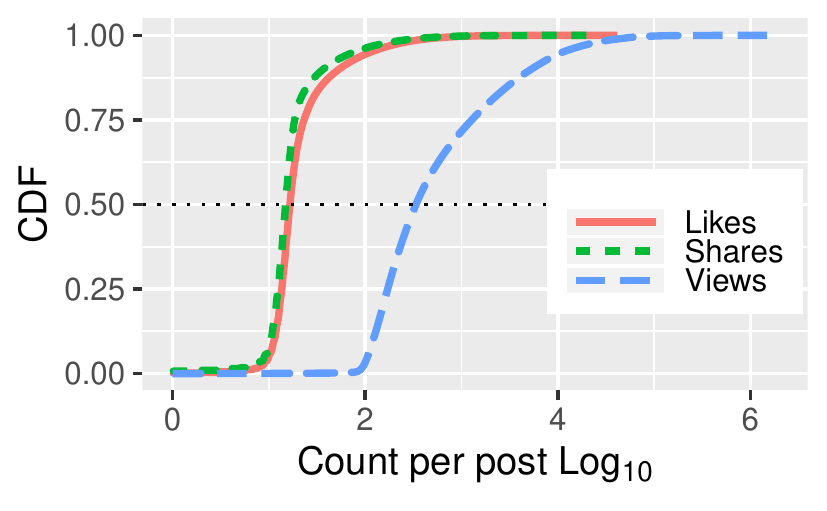}
  \caption{Views, shares and likes across all languages.}
  \label{fig:CDFviews}
\end{figure}

Next, Figure~\ref{fig:CDFviews} presents the empirical distributions of likes, shares and views across all posts in our dataset. Naturally, we see that views dominate with a median of 340 per post. As seen in other social media platforms, we observe a significant skew with the top 10\% of posts accumulating 76\% (2.47 Billion) of all views~\cite{sastry2012tell,cappelletti2012iarank,tyson2015people}. Similar patterns are seen within sharing and liking patterns. Liking is fractionally more popular than sharing, although we note that   ShareChat does not have a retweet-like share option to share content on the platform. Instead,  sharing is a means of \textit{reposting} the content from ShareChat onto  WhatsApp. 

\subsection{Media types}
\begin{figure}
\centering
  \includegraphics[width=\columnwidth]{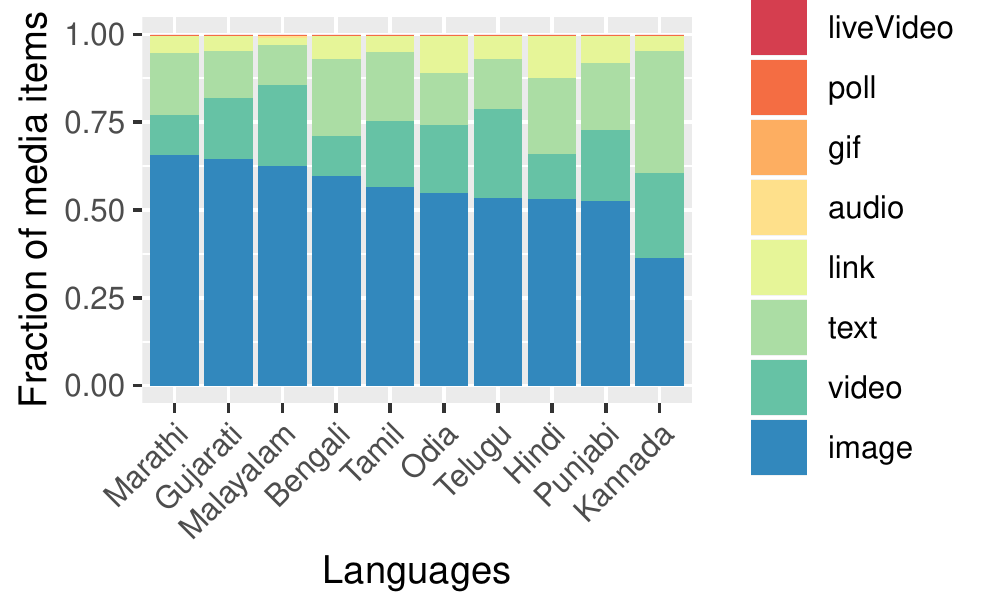}
  \caption{Media count in posts per language. The x-axis is sorted by decreasing count of images in each language.}
  \label{fig:media}
\end{figure}

ShareChat supports various media types including images, audio, video, web links and even survey-style polls. Whereas audio (and audio components of video) would be language specific, images are more portable across languages. Figure~\ref{fig:media} presents the distribution of media types per-post across the different language communities. Overall, images dominate ShareChat with 55\% of all posts containing image content. These are then followed by video (20\%), text (19\%) and links (6\%). It is also interesting to contrast this make-up across the languages. Whereas most exhibit similar trends, we find some noticeable outliers. Kannada (and Haryanvi, not shown) have a substantial number of posts containing web links, as compared to other communities. We conjecture that, overall, the heavy reliance on portable cross-language media such as images and web links may aid the sharing of content across languages.

\subsection{Temporal patterns of activity}
\label{sec:temporal}

\begin{figure}
\centering
  \includegraphics[width=\columnwidth]{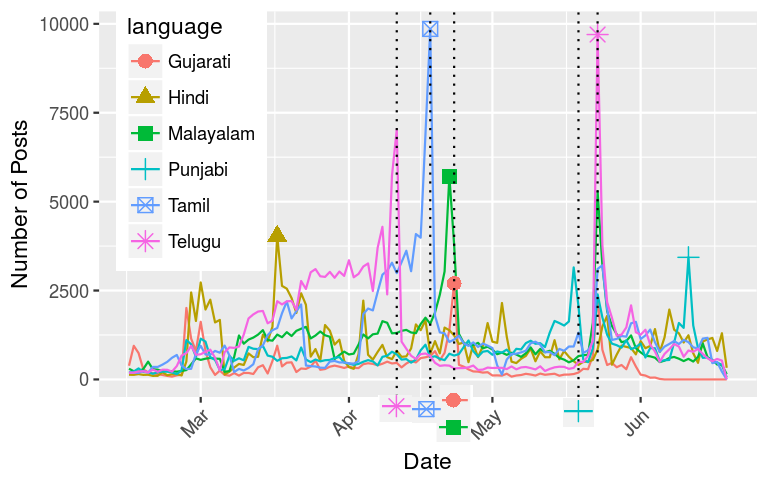}
  \caption{Daily posts plot for languages having maximum peak of more than 2,500 posts per day.  Vertical dotted lines are voting days in one of the phases of the multiple-phase Indian election. The right most vertical dotted line is 23 May, when the results  of the election were announced across the nation, causing peaks in all languages.}
  \label{fig:daily}
\end{figure}

Given that each language appears to primarily have independent content that is native to its language, we next examine how the activity levels of different languages vary across time. Figure~\ref{fig:daily} presents the time series of posts per day across languages. 

Some synchronisation in activity volumes can be observed across languages, which is likely driven by the electoral activities.  Interestingly, however, the  peaks of different languages are out of step with each other in many cases. Digging deeper, we find that this is caused by the multiple phases of voting in the Indian Elections, and the peaks correspond to the voting days in the states where those languages are spoken, as marked with the vertical dotted lines (marker below the x-axis shows which state had a voting day corresponding to that peak). For instance, the Voting day for Andhra Pradesh (major Telugu speaking state) is 11th April; Kerala (Malayalam) on 23th April, Punjab (Punjabi) 19th May; Tamil Nadu (Tamil) 18th April. The final peak on 23 May, (corresponding to the election result declaration day) sees a peak for all the languages.
Although intuitive, we see that these language trends are \emph{not} agnostic to the underlying events important to their communities.

\section{Image Spreading Across Languages}
\label{sec:cross_language}
\label{sec:nonnative}

As noted earlier, ShareChat is organised into separate language communities, and users must choose a language when registering. Thus, each language community forms its own silo. In this section, we explore to what extent information (more specifically image content) crosses from one language silo to another. 

\subsection{Crossing languages is difficult}
\label{subsec:cross_language_difficult}
\begin{figure}
    \centering
    \includegraphics[width=0.9\columnwidth]{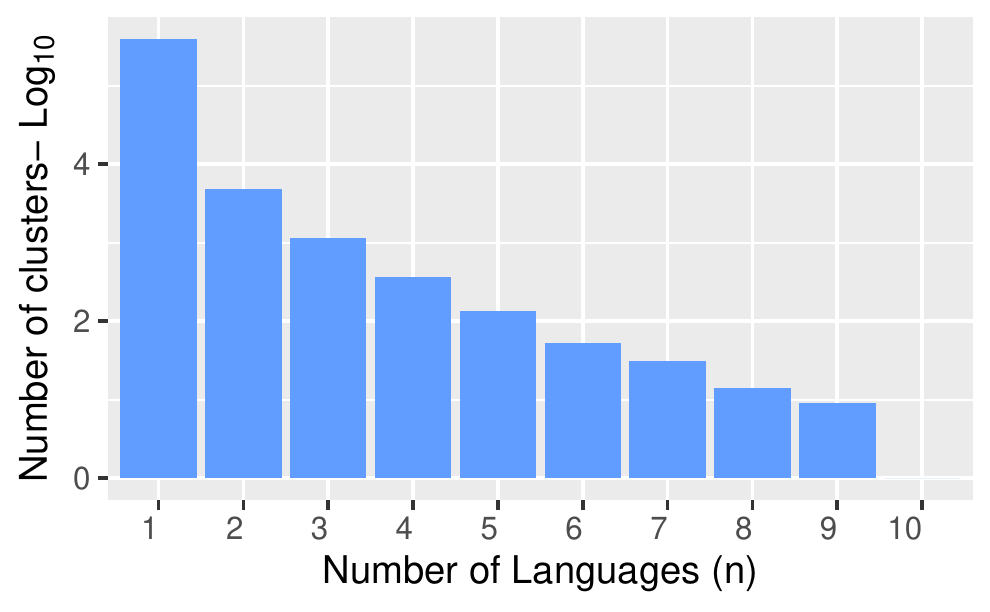}
\caption{Number of image clusters spanning $n$  languages. Each cluster consists of highly similar images as determined by Facebook PDQ algorithm. \emph{y-axis is log scale.}}
    \label{fig:clusters_spanning_lang}
\end{figure}

As noted in \S\ref{sec:img-cluster}, similar images are collected together into clusters using the PDQ algorithm. To test whether users from different languages are exposed to the same image content, we first begin by checking users from how many different languages (as identified by the language set in the profile of the user posting) are represented in each cluster. Figure~\ref{fig:clusters_spanning_lang} presents the number of image clusters that span multiple language communities. 
Note that the y-axis is in log-scale, and the vast majority (98\%) of clusters are images from a single language community. 
However, the remaining 2\% of clusters transcend language boundaries, with 108 clusters (3258 images in total) crossing five different language communities. 
This shows that images \textit{can} be an effective communication mechanism, particularly when contrasted with text (where only 0.3\% of text-based posts occur in multiple language communities using the same methodology).

\begin{figure}[t]
\centering
  \includegraphics[width=\columnwidth]{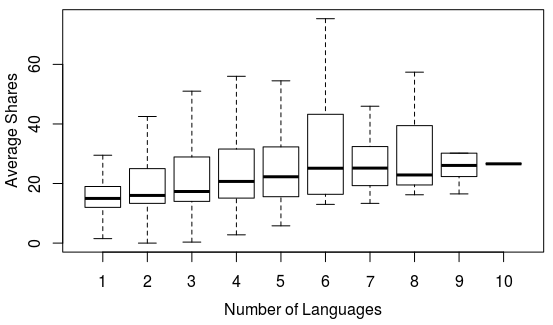}
  \includegraphics[width=\columnwidth]{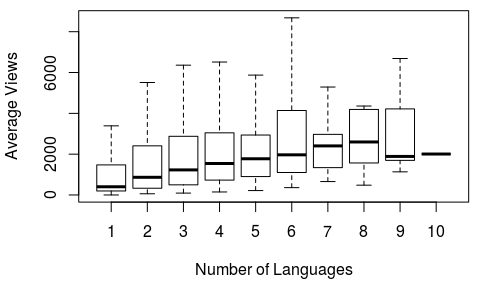}
  \caption{Distribution of the average number of shares (top) and views (bottom) of the image variants represented in a  cluster. We can see that the median increases with the number of languages where images from that cluster are found.}
  \label{fig:shareLang}
\end{figure}

We next test if  images that cross language boundaries proceed to obtain more ``shares'', \ie are they more popular. Recall that shares refer to the act of sharing content via WhatsApp. 
Figure~\ref{fig:shareLang}~(top) presents, as a box plot, the number of shares per item, based on how many language communities it occurs in. 
We see a clear trend, whereby multi-lingual content gains more shares. Whereas content that appears in one language community gains a median of 15 shares, this increase to 20 for 4 languages. This appears to indicate that users may expend more energy to translate content that they find to be `viral' or worthy of sharing widely.
This is intuitive as, naturally, images that move across clusters also gain more views.  Figure~\ref{fig:shareLang}~(bottom) confirms this, showing that the number of views of images in a cluster increases with the number of languages in that cluster. There is a strong (74\%) correlation between number of views on Sharechat and numbers of shares from ShareChat to WhatsApp.

We also noticed the presence of text on the image affects the propensity to share across languages. Recall that 15\% of images do not contain any OCR text. We find that this further impacts sharing rates. The average number of shares for images without text is 22 (standard deviation=74) compared to 34 (standard deviation=155) for images with text. 
A KS test (one-sided) confirms that this difference is statistically significant ($D = 0.09352, p < 0.001$).  
For images with OCR text, 80\% of OCR text is in the same language as the user's profile language. This leaves around 20\% posts that have an OCR text language that is different from the user's profile language.
This is important as it confirms that non-native languages can penetrate these communities. This 20\% is mostly made-up of lingua franca, \ie English (11\% posts, average share 24), Hindi (8\%, average share 34) and Others (1\%).

\subsection{Quantifying cross language interaction}

\begin{figure}[t]
    \centering
 \includegraphics[width=\columnwidth]{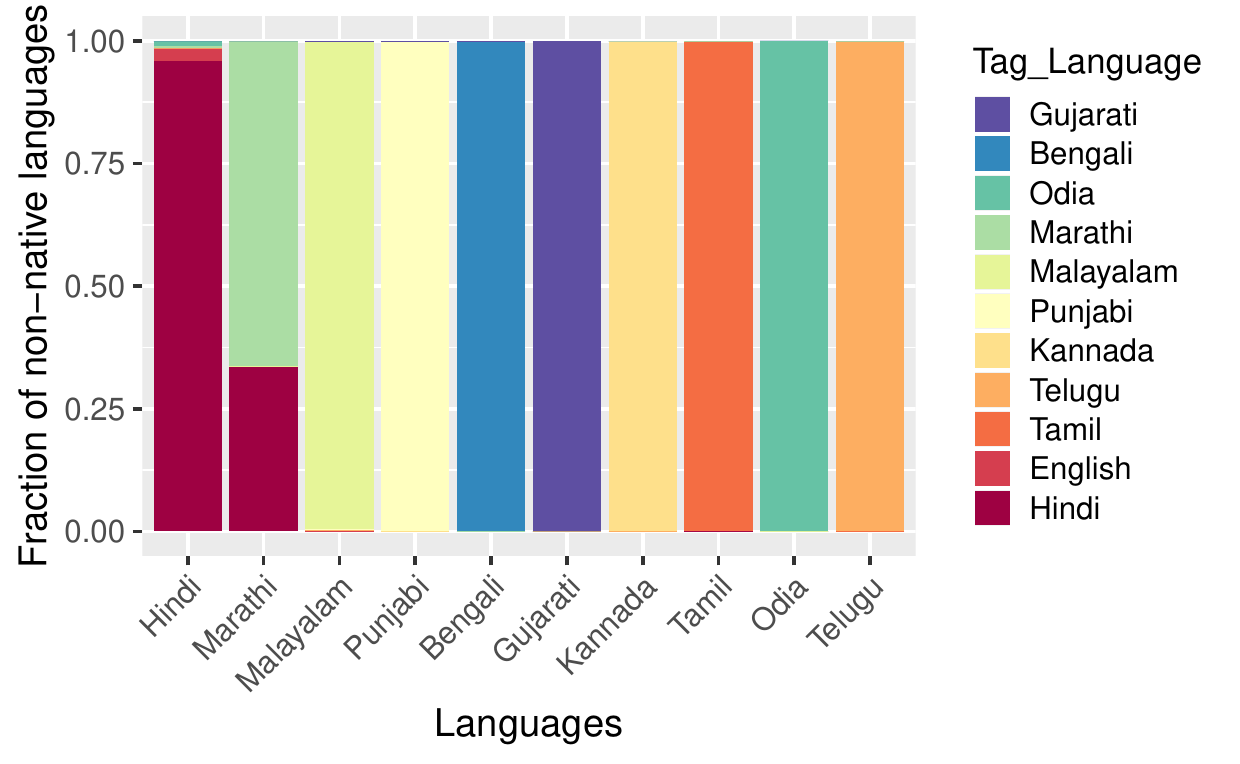}
  \includegraphics[width=\columnwidth]{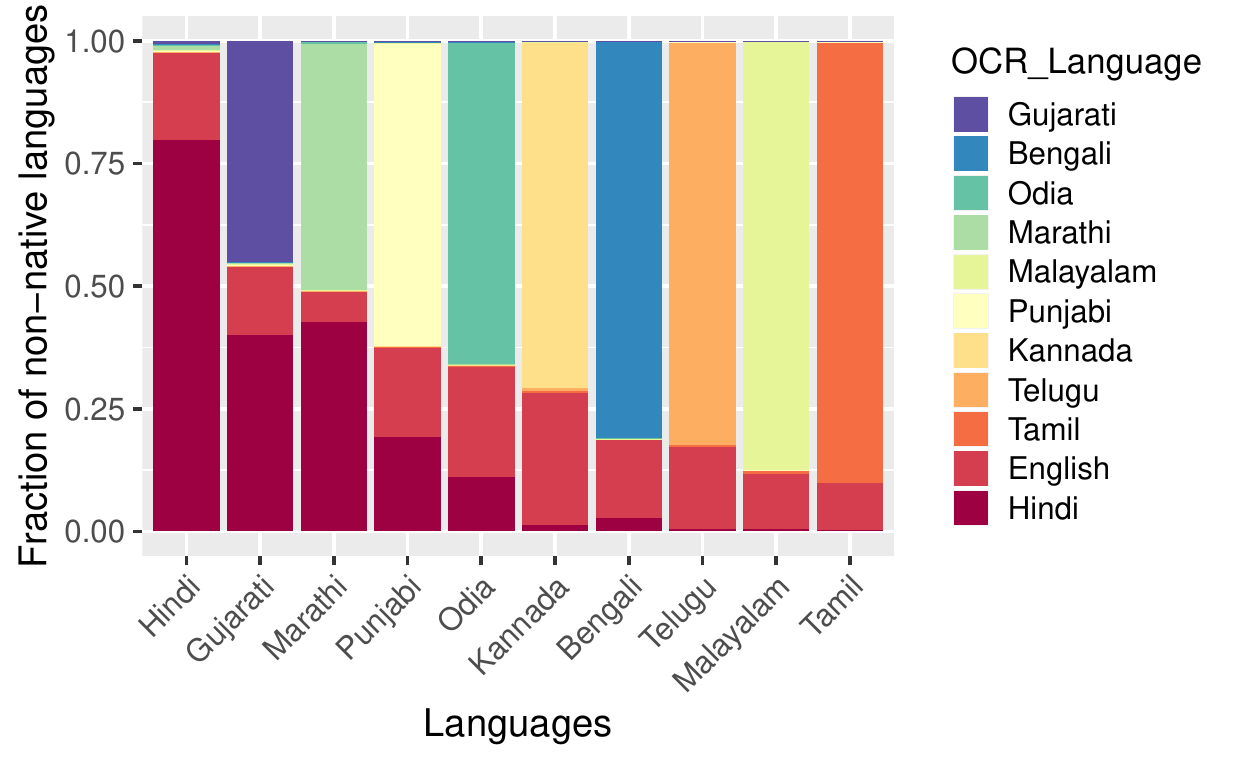}
    \caption{Proportion of non-native languages from (top) hashtags (text)  and (bottom) OCR text (images). Languages on x-axis indicate the user profile language and are ordered by descending proportion of English + Hindi, two languages which are used and understood widely across India. }
    \label{fig:crosslang}
\end{figure}

Based on the above observation, we next take a deeper look at which languages co-occur together and consider cross language interactions via both direct text (hashtags), as well as the text contained within images (\ie OCR text). 
To measure the extent to which some piece of information may go from one language silo to another, we take all posts from users of a specific language, and detect the languages of tags used on those items using Google's Compact Language Detector 2~\cite{cld3}. Similarly, for all images posted by users from a specific language, we detect the language of any OCR text embedded in those images.

Figure~\ref{fig:crosslang}~(top) presents the language make-up for tags within each language community or silo, and Figure~\ref{fig:crosslang}~(bottom) shows the language make-up for the OCR text taken from images. We see that in most language communities, the dominant language for both tags and OCR text tends to be the language of that community. However, we \emph{do} observe a number of cases for languages bleeding across community boundaries. 
Although it is less commonly observed in tags, we regularly see images containing text in other languages shared within communities (as discussed in \S\ref{subsec:cross_language_difficult}). 
As the lingua franca, Hindi and English are by far the most likely OCR-recognised languages to be used in other communities. 
In fact, together Hindi and English make up over half of the image (OCR) text in the Gujarati and Marathi communities. That said, this also extends to other less widely spoken languages too, \eg the Odia community contains noticeable fractions of tags in English and Marathi, as well as Odia itself. 
This confirms that it \emph{is} possible to transcend language barriers, although clearly the prominence of the local language shows that this is not trivial.

\subsection{Drivers of cross-language interaction}
\begin{figure}[t]
\centering
\includegraphics[width=\columnwidth]{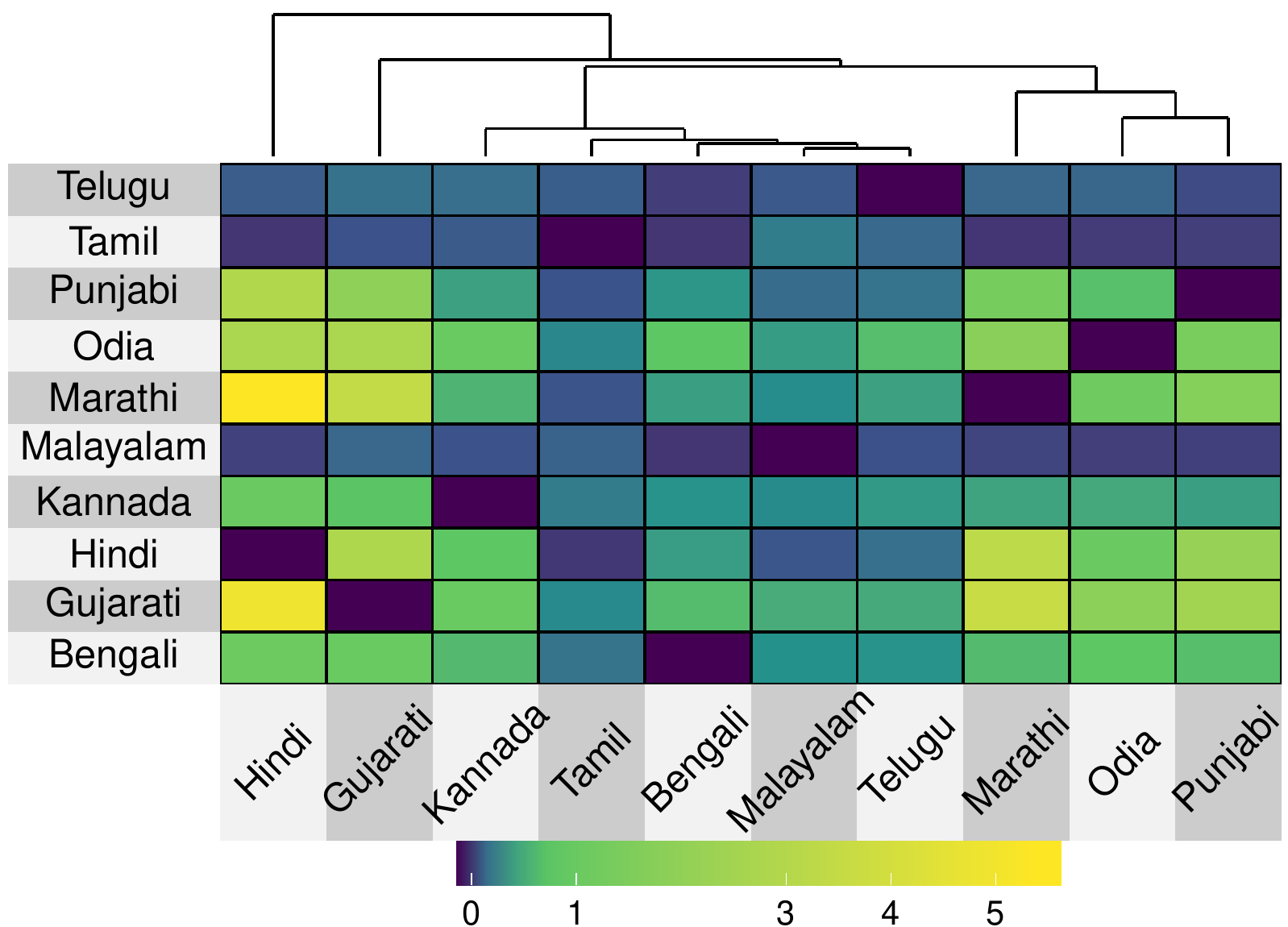}  
\caption{Co-occurrence of languages in image clusters}
  \label{fig:translation-cooccur}
\end{figure}

The previous subsection hints at one possible  factor that may lead to the use of a foreign language -- the widespread understanding of lingua franca such as Hindi and English.
We next explore the extent to which relationships between languages affect whether images are shared across those languages. There are three major language families in India: the Indo-Aryan languages, Dravidian and Munda~\cite{emeneau1956india}, of which the first two are represented in ShareChat.  It is more common to find speakers who can speak both languages of a pair, for languages within the same language family. To get a better understanding of how language communities intersect via image sharing, Figure~\ref{fig:translation-cooccur} presents a dendrogram to highlight how different language communities tend to share material. We  measure the co-occurrence of the same image (or close variants of the same image) in  different language communities. Thus, we measure the extent to which the same or similar information is conveyed in two different languages, for every possible language pair. 

The dendrogram recovers similarities between languages that are geographically or linguistically close to each other (e.g., Dravidian languages such as Tamil, Telugu, Malayalam and Kannada). Table~\ref{table:RL1L2} shows the fraction of posts of a given language $L2$ that occurs within the community of a language $L1$, focusing on the five pairs of languages with the maximum and minimum overlap. Four of the top five overlaps are observed between Hindi and languages of states where Hindi is widely spoken and understood even if it is not the official state language (Gujarat, Punjab and Maharashtra, which speaks Marathi). The only top overlap where Hindi is not involved is between Marathi and Gujarati, languages of neighbouring states, with a long history of interaction and migration. The bottom of the table shows that language pairs with the least cross-over are those from different language families (e.g., Tamil, a Dravidian language and Marathi, an Indo-Aryan language).

\begin{figure}[t]
    \centering
    \includegraphics[width=0.9\columnwidth]{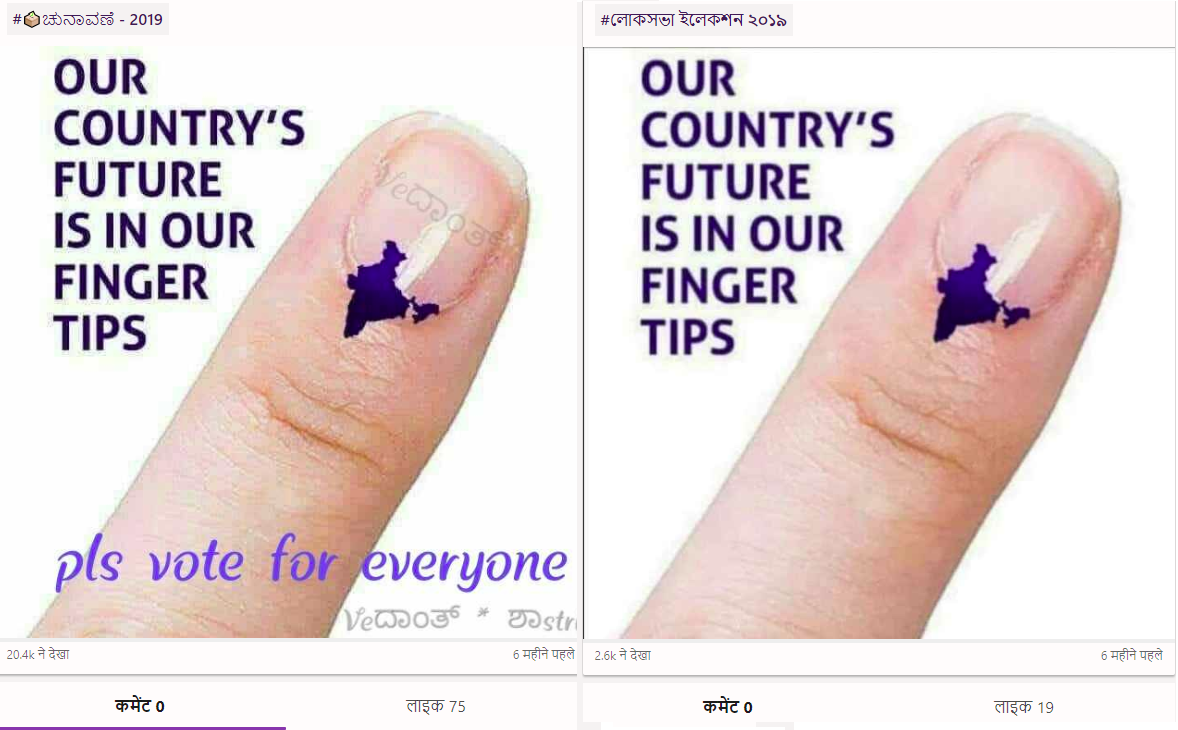}
    \caption{A ``go and vote'' message shared across Kannada (left), Bengali (right), and other languages.}
    \label{fig:go-vote}
\end{figure}

Interestingly, however, the dendrogram also points to close interactions among some pairs of languages that come from very different language families and are spoken in states that are geographically far apart, such as Bengali and Kannada. Manual  examination reveals that many of the posts shared between these two languages are memes containing exhortations to vote (Figure~\ref{fig:go-vote} shows an example).
W.\ Bengal (where Bengali is spoken) and Karanataka (where Kannada is spoken) both went to polls on the same day, which suggests that the shared content between these two languages may have resulted from an organised effort to share election-related information more widely.

\begin{table}[t]
	\caption{Maximum (top 5) and minimum (bottom 5) of overlaps among  pairs of languages}
	\centering
	\begin{tabular}{|p{2.3cm}|p{1.5cm}|p{1.5cm}|p{1.2cm}|}
	\hline
	Rank & $L1$ &$L2$& Overlap\\
	\hline
    1&Marathi&Hindi&5.48\%\\
    2&Gujarati&Hindi&5.01\%\\
    3&Gujarati&Marathi&3.72\%\\
    4&Hindi&Marathi&3.36\%\\
    5&Punjabi&Hindi&3.04\%\\
    …&…&…&…\\
    86&Hindi&Tamil&0.13\%\\
    87&Tamil&Marathi&0.12\%\\
    88&Malayalam&Bengali&0.12\%\\
    89&Tamil&Bengali&0.12\%\\
    90&Tamil&Hindi&0.12\%\\
	\hline
	\end{tabular}
\label{table:RL1L2}
\end{table}

\subsection{What gets translated?}

\begin{figure}[t]
    \centering
    \includegraphics[width=\columnwidth]{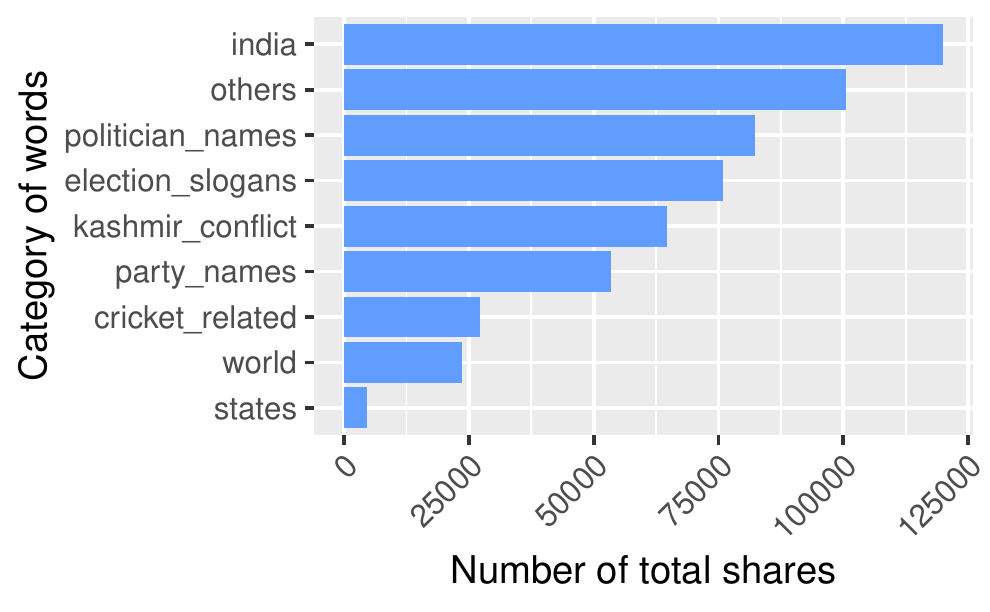}
    \includegraphics[width=\columnwidth]{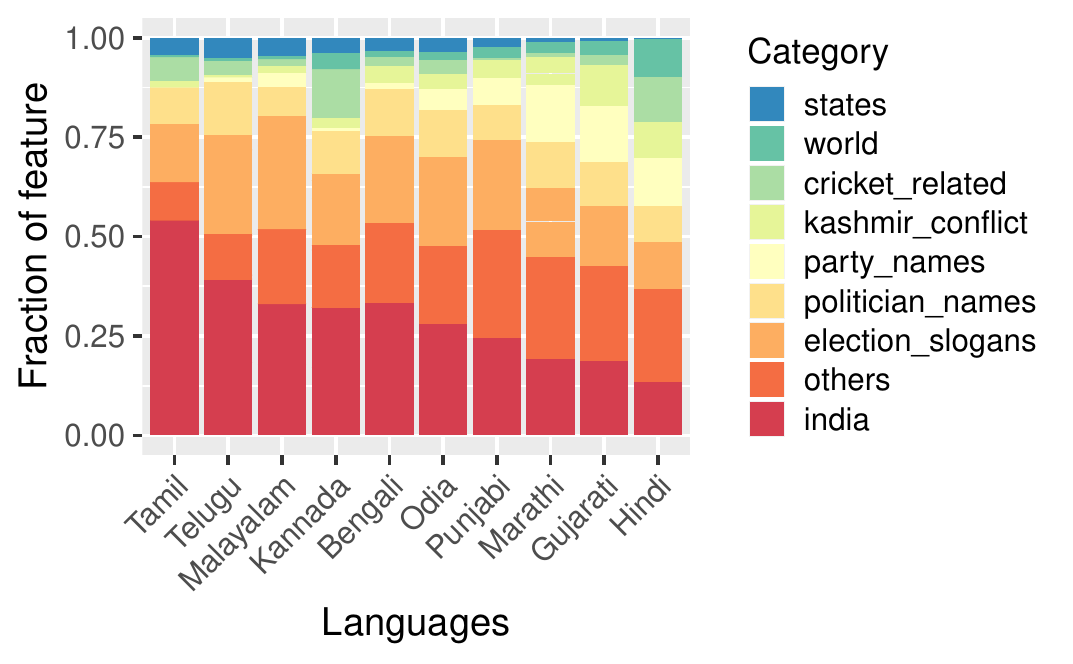}
    \caption{Most popular categories of posts that are shared across language barriers and number of shares (top). Breakdown of topics per language community (bottom).}
    \label{fig:typesofcrosslangcontent}
\end{figure}

We next look at what \textit{kinds} of content get translated and move across languages. To understand this, we first created a list of the most commonly occurring words in the OCR text of the images (See \S\ref{sec:ocr}). 
Taking all the words which occur more than five times into consideration, we manually code them into 9 categories. We follow a two step approach to come up with this categorisation. First, different authors of the paper coded a small subset of images to come up with a coarse set of categories. These were merged to come up with the final list of 9 categories of information which transcend language boundaries: election slogans, party names,  politician names, state names, Kashmir conflict, cricket, India, world and
others (such as viral, years, family, share and so forth).
Figure~\ref{fig:typesofcrosslangcontent}~(top) shows the total number of shares that each of these categories get. This reveals an interesting mix of topics that provide new insights: since we collected data related to politics, keywords related to politics such as politician/party names, \etc are expected. However, it is interesting to see categories related to issues such as the Kashmir conflict and cricket --- these topics are of interest across the nation; possibly an important consideration when deciding which images to translate across languages. 

Briefly, Figure~\ref{fig:typesofcrosslangcontent} (bottom) also highlights the presence of these topics across the language communities. We see that trends are broadly similar, yet there are noticeable differences. For instance, ``India'' makes up over half of the posts in Tamil compared to less than 10\% in Hindi. Individual language communities also show different preferences; for instance, Malayalam shares the highest proportion of election slogans. Similarly, Hindi, Marathi and Gujarati share more party names-related images than other communities, whereas Hindi and Kannada share more cricket images than other languages. Given that Indian states were created based on the languages spoken, cross-language sharing of state-specific issues are negligibly small in all communities.

\subsection{Are translations faithful?}
Our previous results have shown that meme-based images (containing text) that transcend language barriers often benefit from translation. Due to this, we are interested to know whether such translations preserve or alter the semantic meanings. Since many of the most widely translated categories are related to politics or contentious national issues in India, this question acquires an additional dimension of truth and propaganda.

\begin{figure}
    \centering
    \includegraphics[width=0.9\columnwidth]{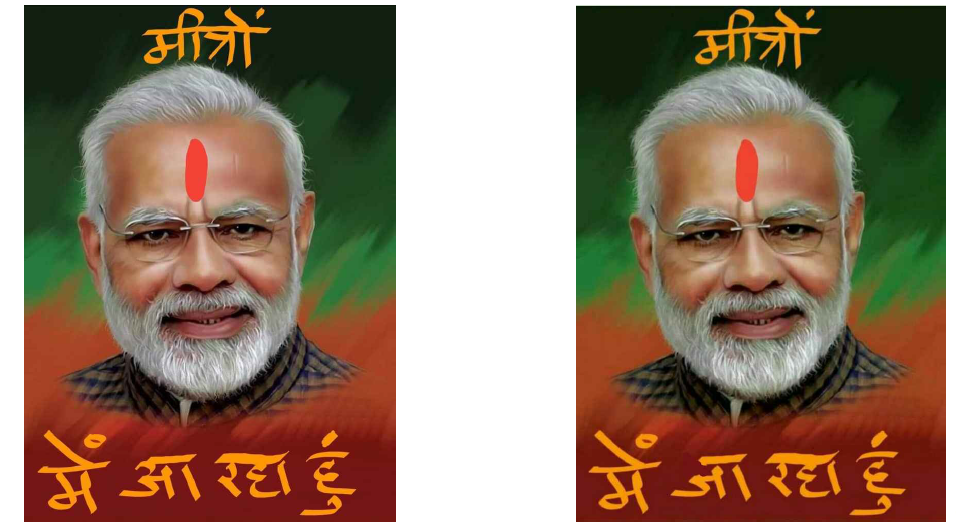}
    \caption{Example of images with different messages which portray a political message with different messages in Hindi. Both show Mr. Modi, the incumbent Prime Minister of India at the time of the Election. The caption of the left figure reads ``\emph{Friends, I am coming}'', whereas the right figure reads ``\emph{Friends, I am going}''.
    }
    \label{fig:meaningTwo}
\end{figure}

We have 68 image clusters with multiple languages, consisting of 1080 images. Of these, we are able to fully translate all images within 63 of those clusters.\footnote{The other clusters contain Odiya or Punjabi -- languages for which we did not have access to native speakers}
These are made-up of 769 images.
To compare the meanings of each piece of text within an image cluster, we allocate each of the images to 2 annotators. Each annotator looks at all of the images in a cluster, as well as the associated OCR text translated into English.\footnote{The translations are high quality manual translations by native speakers as described in \S\ref{sec:ocr}}
The annotators are then responsible for grouping all images within the cluster into semantic groups, based on their OCR texts.
This may yield, for example, a single image with two different OCR texts with diametrically opposite meanings (e.g., see Figure~\ref{fig:meaningTwo}). We then say that there are two different ``messages'' contained in these clusters. 

\begin{figure}[t]
\centering
 \includegraphics[width=\columnwidth]{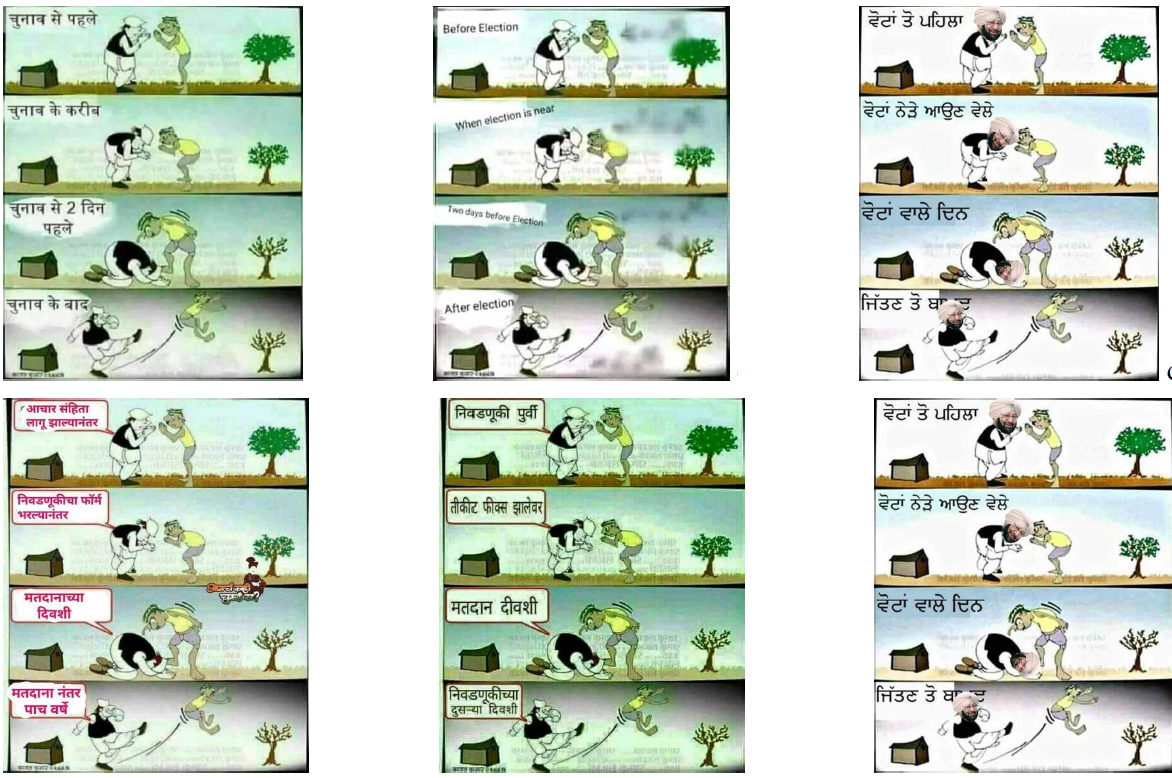}
 \caption{Meme of a similar post in seven different languages (only a selection of languages shown) by nine different language communities in 190 images. Each variant is a joke about how politicians court citizens, becoming very courteous just before the elections (politician bowing down) but then turn on them after getting elected (bottom image: politician kicking the citizen).}
 \label{fig:pdqMeme}
\end{figure}

\begin{figure}[t]
    \centering
    \includegraphics[width=0.9\columnwidth]{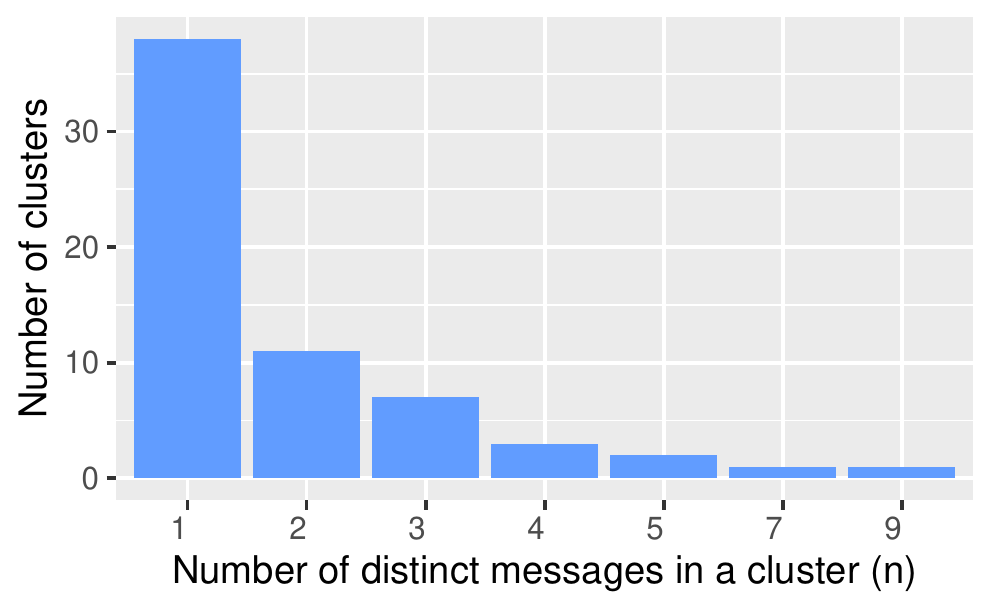}
    \caption{Number of image clusters where OCR texts contain more than $n$ distinct messages.}
    \label{fig:MeaningNumber}
\end{figure}

We find that images contained in the majority of clusters have similar messages even when the  text is translated into multiple languages (Figure~\ref{fig:pdqMeme} shows an example). However, we find a handful of clusters with more than one message: 11 have two messages; often these are memes, but with the ``translations'' containing distinct messages that have different meanings. Figure~\ref{fig:MeaningNumber} shows that in all, only 25 clusters have more than one message, and the number of distinct messages in each cluster tends to be small: only 2 clusters have more than 5 messages, and one cluster has 9 different messages. 
Interestingly, we find that image clusters containing more than one message tends to get more shares (mean 44, median 24) and views (mean 2865, median 1889) than clusters where the  images contain only one meaning (mean shares 26, median 20; mean views 2255, median 1329).

\begin{figure}[t]
    \centering
    \includegraphics[width=\columnwidth]{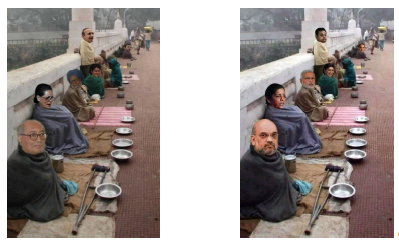}
    \caption{Example of non-text cluster where political faces are changing. Both images depict politicians as beggars. The left image shows leading politicians from the \textit{opposition} party and the right image replaces those heads with leaders of the \textit{ruling} party.}
    \label{fig:noOCRmeme}
\end{figure}

Finally, we briefly note that although  our detailed manual coding and analysis supported by native speakers (\S\ref{sec:ocr}) suggests that most of the cases where images have different messages is caused by differences in the \emph{text} embedded in the images, we have also observed a few cases where the images themselves have been changed, creating a new meaning. Figure~\ref{fig:noOCRmeme} shows an example.

\section{Conclusions}

This paper investigated the role of language in the dissemination of political content within the Indian context, through the lens of ShareChat. 
We began by asking two sets of research questions. 

First, \emph{can image content transcend language silos and, if so, how often does this happen and amongst which languages?} We find that the vast majority of image clusters remain within a single language silo. However, this leaves in excess of 33k images crossing boundaries. We find that geography and language connections play a role: when content crosses language boundaries, it does so into languages which belong to neighbouring states, or which are related linguistically.

Second, we asked \emph{what kinds of images are most successful in transcending language silos, and do their semantic meanings mutate?}
We certainly observe certain topics that are more effective at gaining cross language interest. As anticipated, we find that images containing text related to national Indian politics cross languages more often. But we further observe other topics of pan-Indian interest (\eg~cricket) also gain traction among a diverse set of languages. By clustering images based on perceptual similarity, and manually verifying their semantic meanings with annotators, we found that 25 clusters of images had text which changed meaning as they crossed language barriers.

There are a number of lines of future work and, indeed, this paper constitutes just our first steps. We plan to focus more closely on the semantic nature of images, including the characteristics that lead to more popular posts, as well as posts that can overcome language barriers. We also plan to understand other social broadcasts such as live videos, poll and links in these user posts~\cite{raman2018facebook}. Preliminary analysis of the ShareChat content has revealed the presence of ``fake news'', and has shown how it tends to gain higher share rates than mainstream content. Thus, another line of investigation is to trace the origins of such content and understand how it may link to more targeted political activities~\cite{bhatt2018illuminating,agarwal2020stop,starbird2017examining}.

\section{Acknowledgements}
We sincerely thank people and funding agencies for assisting in this research. We acknowledgement the support given by Aravindh Raman, Meet Mehta, and Shounak Set from King's College London, Nirmal Sivaraman from LNMIIT for help with translations and Tarun Chitta for help with collecting the data. We also acknowledge support via EPSRC Grant Ref: EP/T001569/1 for ``Artificial Intelligence for Science, Engineering, Health and
Government'', and particularly the ``Tools, Practices and Systems'' theme for ``Detecting and Understanding Harmful Content Online: A Metatool Approach'', King’s India Scholarship 2015 and a Professor Sir Richard Trainor Scholarship 2017. The paper reflects only the authors' views and the Agency and the Commission are not responsible for any use that may be made of the information it contains.

\bibliography{FULL-AgarwalP.1212}
\bibliographystyle{aaai}

\end{document}